\documentclass{nature1}
\usepackage{color}
\usepackage[pdftex]{graphicx} %for embedding images
\usepackage{subfigure} %(a)(b)(c) figures
\usepackage{float}% figures closely follow paragraph

\title{Observation of Quantum Zeno Blockade on Chip}

\author{Jia-Yang Chen$^{1,2}$, Yong Meng Sua$^{1,2}$, Zi-Tong Zhao$^{1}$, Mo Li $^{1,2}$, and Yu-Ping Huang$^{1,2}$}

\begin{document}

\maketitle

\begin{affiliations}
 \item Department of Physics and Engineering Physics, Stevens Institute of Technology, Hoboken, New Jersey 07030, USA
 \item Center for Distributed Quantum Computing, Stevens Institute of Technology, Hoboken, New Jersey 07030, USA
\end{affiliations}

%%%%%%%%%%%%%%%%%%%%%%%%%%%%%%%%%%%%%%%%%%%%%%%%%%%%%%%%%%%%%%%%%%%%%%
\begin{abstract}
When overlapping in an optical medium with nonlinear susceptibility, light waves can interact with each other, changing their phases, wavelengths, shapes, and so on. Such nonlinear effects, discovered over a half century ago, have given rise to a breadth of important applications. Applying to quantum-mechanical signals, however, they face fundamental challenges arising from the multimode nature of the interacting electromagnetic fields, such as phase noises and Raman scattering. Quantum Zeno blockade allows strong interaction of light waves without them physically overlapping, thus providing a viable solution for those challenges, as indicated in recent bulk-optics experiments. Here, we report on the observation of quantum Zeno blockade on chip, where a light wave is modulated by another in a distinct ``interaction-free'' manner. For quantum applications, we also verify its operations on a single-photon level. Our results promise a scalable platform for overcoming several grand challenges faced by nonlinear optics and quantum information processing, enabling, e.g., manipulation and interaction of quantum signals without decoherence.
\end{abstract}
%%%%%%%%%%%%%%%%%%%%%%%%%%%%%%%%%%%%%%%%%%%%%%%%%%%%%%%%%%%%%%

The quantum Zeno effect occurs when a coherently evolving quantum system is strongly coupled to external degrees of freedom, often referred to as ``environment'' or a reservoir, with the result that the evolution is suppressed or frozen \cite{misra1977zeno,peres1980zeno,joos1984continuous}. A counter-intuitive conception stemming from a philosophical argument over 2000 years ago, namely the ``Achilles and the Tortoise'' paradox, the Zeno effect is in fact implied in the fundamental postulates of quantum mechanics, where measurement causes wavefunctions to collapse. Since its first observation in a trapped-ion system in 1990 \cite{ItaHeiBol1990PRA}, the Zeno effect has been exploited for ``interaction-free'' measurement  \cite{elitzur1993quantum,kwiat1995interaction,kwiat1996quantum,tsegaye1998efficient,kwiat1999high}, by which a ``bomb'' can be probed by a photon without the two to exchange any energy quantum, i.e., physical interaction. First appearing as an attractive topic of fundamental interests, it has later been studied for fascinating applications in optics, atomic physics, and quantum information, such as counterfactual quantum computing \cite{hosten2006counterfactual}.

Exploiting the quantum Zeno effect in nonlinear optics, we have analyzed \cite{huang2010interaction1,huang2010interaction2,huang2012interaction} and verified in two independent experiments the notion of quantum Zeno blockade (QZB) \cite{mccusker2013experimental,strekalov2014progress}. QZB occurs when a photonic system interacts with external degrees of freedom through a nonlinear-optical channel, so that when the nonlinear coupling is strong, occupation of a certain mode of the system by a single photon can ``block'' (more precisely, suppress) additional photons from coupling into the system. Here, the nonlinear interaction functions as a continuous probe ``monitoring'' the system's photon occupation, thus freezing its additionally populating through the quantum Zeno effect. The interaction can be dissipative, like that of two-photon absorption \cite{jacobs2009all,huang2012antibunched}, or be coherent such as sum or difference frequency generation \cite{huang2010interaction2}.

One fascinating aspect of QZB, among others, is that it enables strong interaction between optical waves without them physically overlapping \cite{huang2010interaction1,huang2010interaction2,huang2012interaction}. For example, using second-order nonlinear Fabry-Perot and whispering-gallery-mode (WGM) cavities, we have observed  interaction-free all-optical switching, where a signal field is switched by a pump field only due to a \textit{potential} for nonlinear coupling between the two, but without such coupling actually happening in the asymptotic limit \cite{mccusker2013experimental,strekalov2014progress}. Notably, unlike the photon blockade mediated by single atoms, here no excitation, either excited atoms, excitons or their equivalences, is physically created in order for the blockade to take effect \cite{birnbaum2005photon}. This distinct interaction-free and excitation-free features combined eliminate the phase noise, dissipation, and decoherence (such as via spontaneous emission) \cite{huang2010interaction2,huang2012interaction}, thus overcoming grand challenges for all-optical processing of quantum or photon-level classical signals \cite{shapiro2006single,chang2014quantum}. Recently, we have found that implemented in a low-loss nonlinear microcavity, QZB allows deterministic interaction between single photons, which may enable scalable photonic quantum computing, i.e., without post selection nor use of auxiliary photons \cite{sun2013photonic}.

Despite those great promises, to date all demonstrations of quantum Zeno effects have used bulk optical systems, which are not suitable for scaling \cite{jacobs2009all,mccusker2013experimental,strekalov2014progress}. Here we demonstrate quantum Zeno blockade on a scalable, chip-integrated platform, which has shown superior performance for nonlinear and quantum optics \cite{fortsch2013versatile,SecondGaAs}. Using WGM nanocavities fabricated on lithium niobate on insulator (LNOI), we have observed strong Zeno blockade for interaction-free and excitation-free all-optical operations. As a validation of our platform for quantum applications, we have also measured the noise level due to spontaneous photon scattering, and demonstrated quantum Zeno modulation of single photons. Our experiments mark a significant step toward scalable quantum information processing based on deterministic and noise-free logical operations for single photons.

The rest of this manuscript is structured as follows. In Section 1, we describe our LNOI nanocavity and characterize its nonlinear response with second-harmonic generation. In Section 2, we present the Zeno blockade results for both continuous-wave and pulsed optical signals. In Section 3, we assess the background noise level of the current nanocavity and demonstrate interaction-free and excitation-free modulation of single photons. Lastly, we conclude in Section 4.

\section{LNOI nanocavity and second harmonic generation}
The quantum Zeno blockade exploits the strong second-order nonlinear ($\chi^{(2)}$) effects offered by an air-suspended microdisk cavity nanofabricated on a z-cut LNOI, see Fig.~1(a). The fabrication procedure is described in the Methods Section. For this application, a major challenge is to attain phase matching for the disparate wavelengths across the telecom C-band and near-IR band. To this end, we utilize both the strong birefringence of the lithium niobate material and the geometric dispersion of the strongly confined WGM's in the LNOI microdisk to offset the large chromatic dispersion over an octave. For the former, we use quasi-transverse-electric (quasi-TE) modes for the telecom light waves and quasi-transverse-magnetic (quasi-TM) modes for the near-IR wave. For the latter, we start with a 500-nm thick LNOI thin film and etch it down to only 200 nm using an inductively coupled plasma (ICP) system. By such, we achieve phase matching for the second-harmonic generation between light waves near 1550 nm and 775 nm, all in the fundamental spatial modes along the cross section. Figure 1(b) shows an example of such phase-matched modes based on a finite element method (FEM) analysis, which is verified in our experiment. The fact that the two modes correspond to alike optical fields, each strongly confined to sub-wavelength and singly peaked, leads to strong mode overlapping and thus offers a superior nonlinear coupling efficiency between the modes \cite{moore2016efficient}. The disadvantage with the current use of the birefringence, however, is that for the z-cut crystal orientation, the $\chi^{(2)}$ processes are through the $d_{31}$ and $d_{32}$ tensor elements, which are 4 to 5 times lower than the largest element $d_{33}$ of lithium niobate. We expect to overcome this deficiency by applying centric periodic poling \cite{meisenheimer2015broadband} to the WGM nanocavity in the future work. Finally, to spectrally align the cavity modes, we choose the microdisk radius to be around 20 $\mu$m.

\begin{figure}[H]
  \centering
  \subfigure[]{
   \label{fig:subfig1.1.1:a} %% label for 1 subfigure
    \includegraphics[width=2.5in]{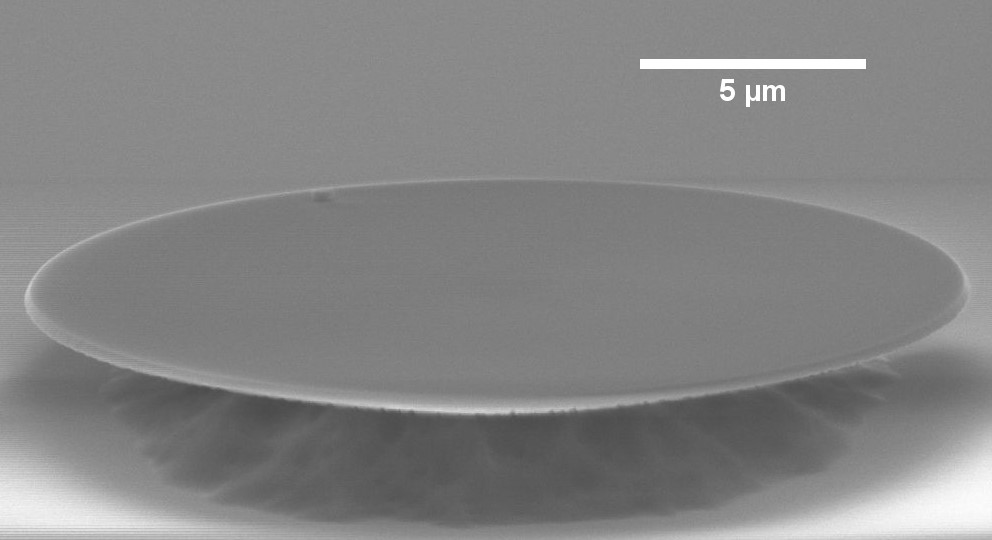}}
  \hspace{0.5in}
  \subfigure[]{
   \label{fig:subfig1.1.1:b} %% label for2 subfigure
    \includegraphics[width=2in]{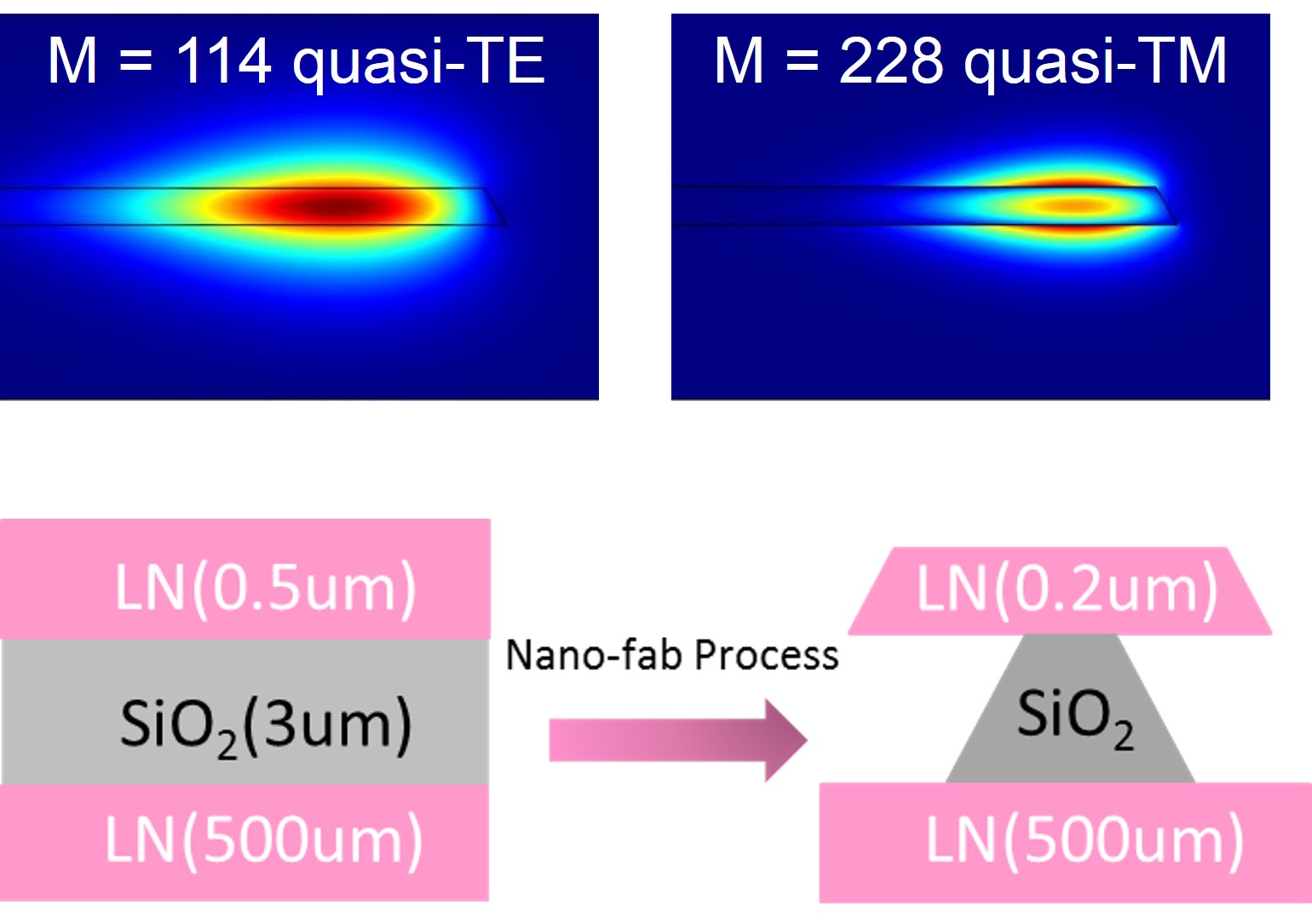}}

  \caption{(a) Scanning electron microscopy (SEM) image of a LNOI microdisk. (b) Simulated profiles of phase matched cavity modes and fabrication schematic of the LNOI microdisk. M denotes the mode's azimuthal order. The quasi-TE and quasi-TM modes are horizontally and vertically polarized relative to the microdisk plane, respectively, near 1550 nm and 775 nm.}
  \label{fig:subfig1.1.1} %% label for entire figure
\end{figure}

For the on/off chip light coupling, we use a piece of tapered fiber as an evanescent coupler for the microdisk, as shown in Fig. 2(b). The fiber is tapered down to a sub-micron diameter from a piece of standard single-mode fiber (SMF-28) using a home-built fiber-tapering system incorporating flame brushing \cite{harun2013theoretical}. Its transmission loss can be as low as 0.05 dB upon tapering, but increases to around 1 dB after extended exposure in ambient air. The coupling efficiency of the tapered fiber and the microdisk is sensitive to the relative position between them, so that we use a three-axis piezo positioning system to achieve a submicron tuning resolution.

To characterize the cavity resonances, we use an optical spectrum analyzer (OSA) to measure the transmission of a broadband light generated via the amplified spontaneous emission (ASE) of an erbium-doped fiber amplifier (EDFA). A typical spectrum for a 20-$\mu$m-radius microdisk is shown in Fig. 2(a), where the free spectral range (FSR) between the fundamental WGM's is measured to be 9.1 nm near 1546 nm. The smallest cavity linewidth is about 0.06 nm, which corresponds to a loaded cavity quality factor of $2.6\times10{^4}$, as shown in Fig. 2(b).

\begin{figure}[H]
  \centering
  \subfigure[]{
   \label{fig:subfig1.1.1:a} %% label for 1 subfigure
    \includegraphics[width=2.5in]{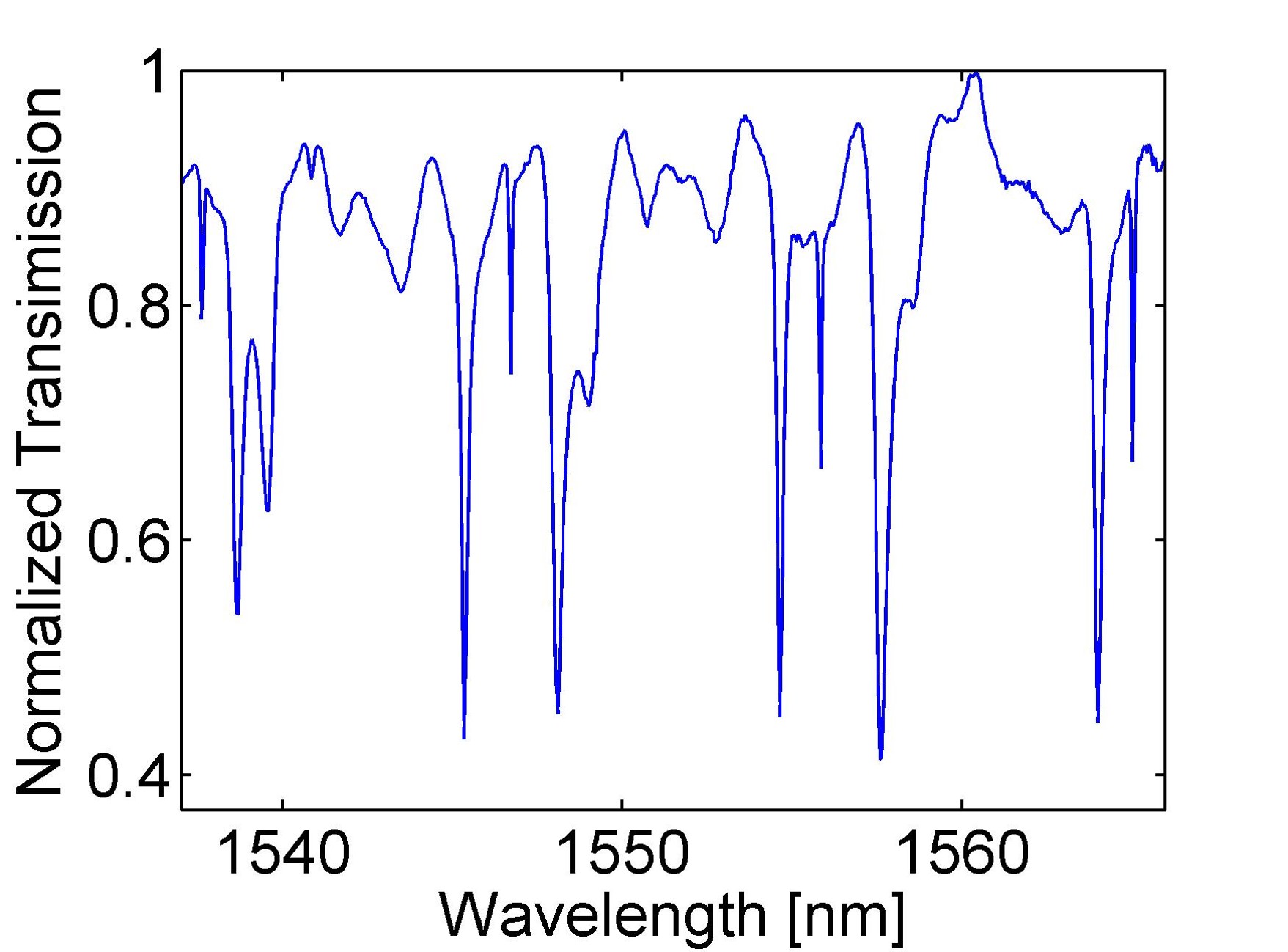}}
  \hspace{0.5in}
  \subfigure[]{
   \label{fig:subfig1.1.1:b} %% label for2 subfigure
    \includegraphics[width=2.5in]{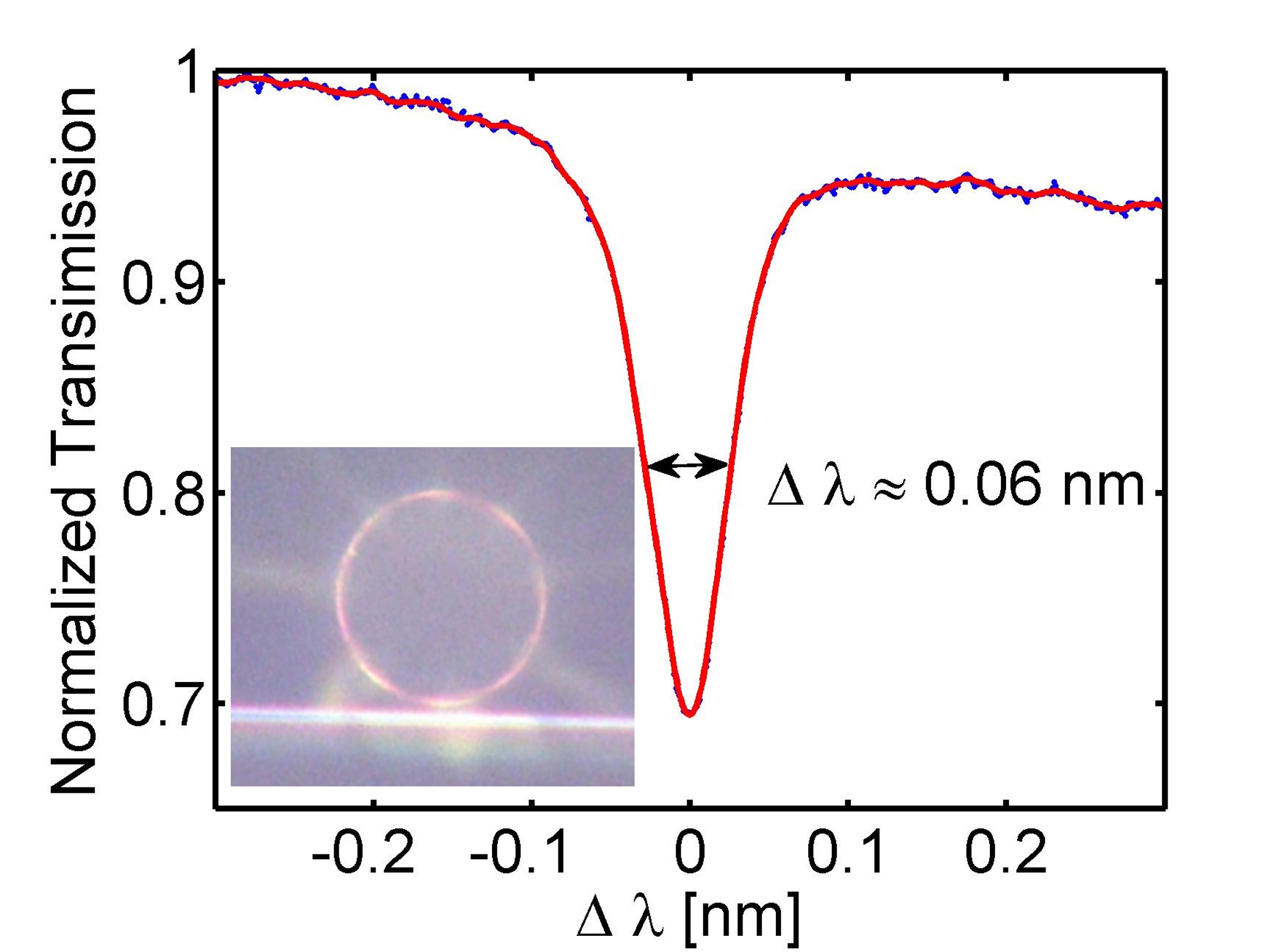}}

  \caption{(a) The transmission spectrum of a LNOI microdisk. (b)The spectrum of one mode which yields
$2.6\times10{^4}$ quality factor at around 1546 nm. The inset is the microscope image of the LNOI microdisk coupled with a tapered fiber.}
  \label{fig:subfig1.1.1} %% label for entire figure
\end{figure}

To examine the phase matching, we measure the second-harmonic generation (SHG) in the microdisk by coupling a miliwatt continuous-wave (CW) pump light into a fundamental quasi-TE mode around 1546 nm. The generated second-harmonic light (around 773 nm) is coupled out of the microdisk through the same tapered fiber and measured by using OSA. The SHG power as a function of the input pump power is shown in Fig. 3, where a clear quadratic dependency is seen, thus verifying the frequency doubling process. Also shown in the figure inset is a micrograph of the generated SH light scattered off the microdisk.

\begin{figure}
\includegraphics[width=3in]{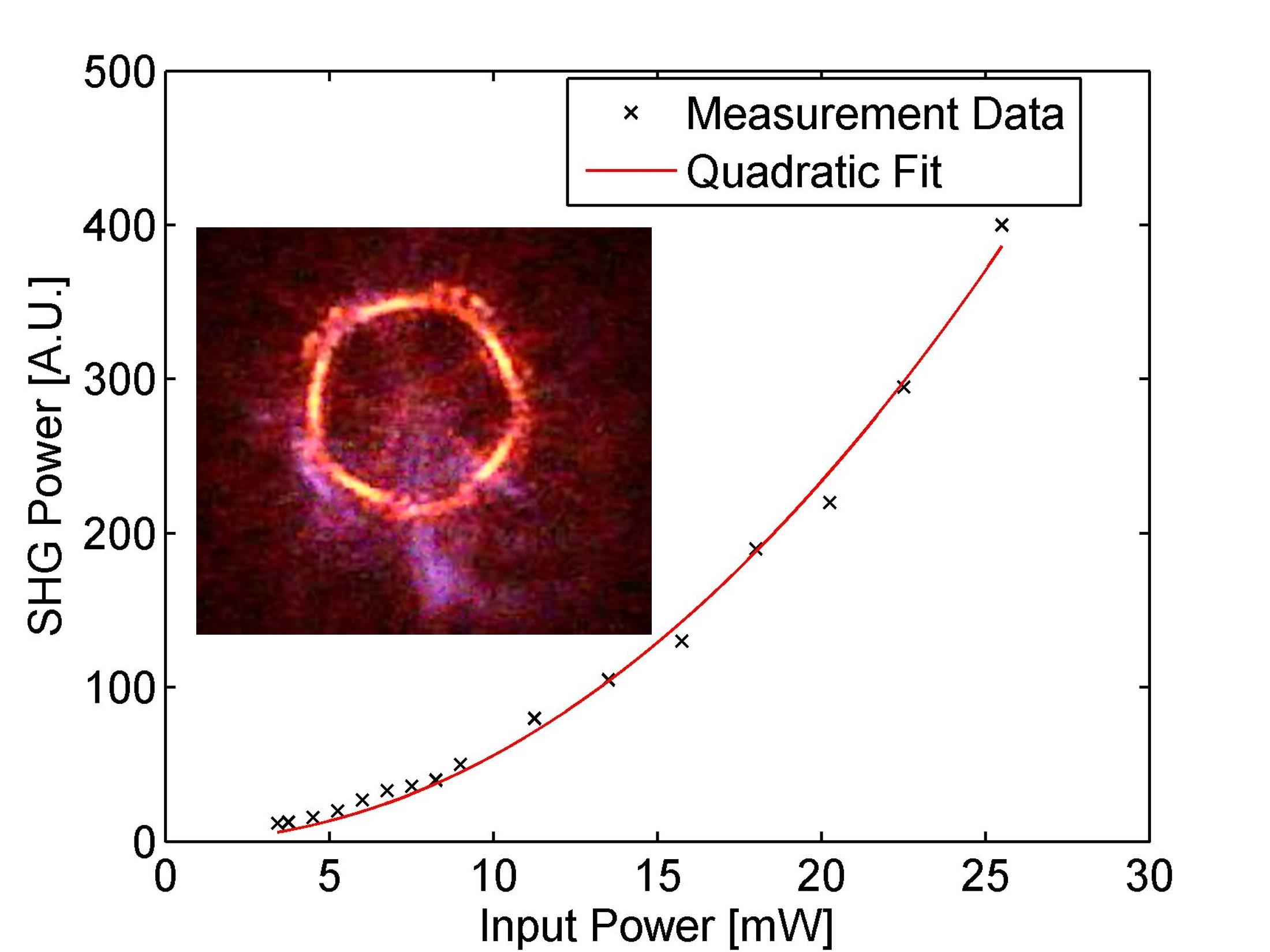} % this command will be ignored
  \centering
\caption{Generated second-harmonic power as a function of input pump power. Inset is the microscope image of the microdisk illuminated by the generated SH light.}
\end{figure}

\section{Experimental setup and Quantum Zeno Blockade}
Once the SHG is optimized, we switch to sum-frequency generation (SFG) between a CW pump and a CW signal wave to further optimize phase matching. It turns out that the strongest SFG occurs between two adjacent fundamental quasi-TE modes, each at 1536.790 nm and 1545.900 nm. For these CW light waves, however, the cavity resonances can thermally drift by as high as 0.1 nm, which exceeds their linewidth. In order to mitigate this thermal effect, we develop an optimization method for stabilizing the SFG; see the Methods Section.

\begin{figure}
\includegraphics[width=5in]{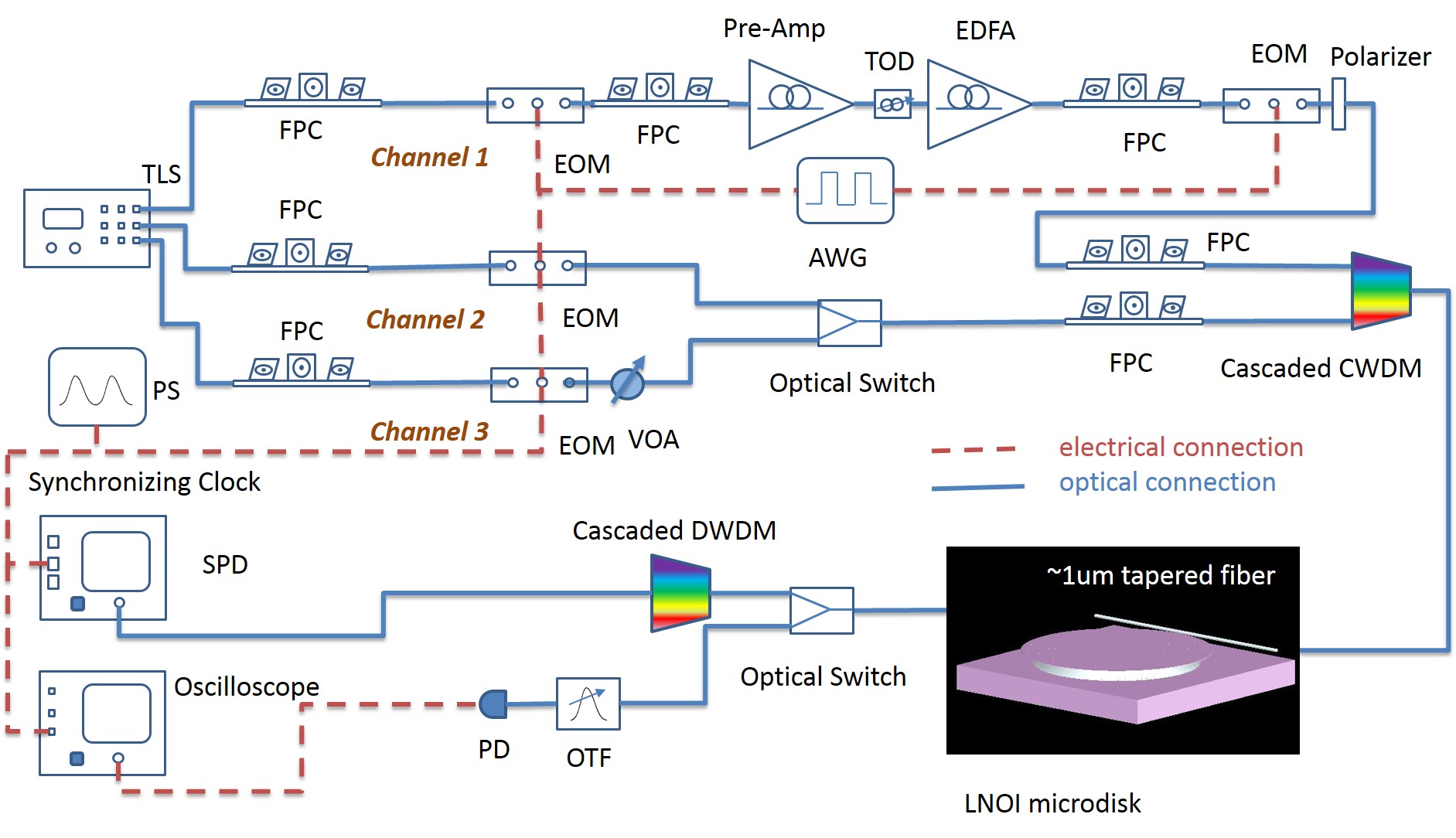} % this command will be ignored
  \centering
\caption{Nonlinear optics experimental setup with three synchronized optical channels. Channel 1 creates a pump pulse train of 1 MHz repetition rate and 250 ps FWHM. Channel 2 produces a quasi-CW signal with 1-MHz repetition rate and 10-ns FWHM. Channel 3 generates signal pulses at a 50 MHz repetition rate and 200 ps FWHM. TLS, multichannel narrow linewidth ($<$100 KHz) tunable laser system, PS, electrical pulse generation system, EOM, electro-optic modulator, FPC, fiber polarization controller, VOA, variable optical attenuator, AWG, arbitrary waveform generator, CWDM, coarse wavelength-division multiplexer, DWDM, dense wavelength-division multiplexer, TOD, tunable optical delay, OTF, optical tunable filter, SPD, single photon detector.}
\end{figure}

Once efficient SFG is observed and thermally stabilized, we quickly swap in a quasi-CW signal and a pulsed pump at the same wavelengths, in order to obtain a higher peak power for strong SFG while maintaining a low average power to be around 250 $\mu$W for minimizing the thermal effect. The experimental setup is shown in Fig. 4. The pump is a 1-MHz pulse train with 250-ps full width at half maximum (FWHM), created by modulating a CW laser using two consecutive electro-optical modulators (EOM's). To achieve the high pump peak power ($\sim$ 1 W), the output of the first EOM is amplified in a two-stage EDFA system. The amplified pulses are then picked by the second EOM to further reduce the repetition rate thus the pump average power. The quasi-CW signal is synchronized and temporally aligned with the pump but a much wider pulse width (about 10 ns FWHM). The resulting pump and signal waves then each pass through a fiber polarization controller, before combined into the microdisk using a pair of cascaded coarse wavelength-division multiplexers (CWDM's).

The average (peak) power coupled into the microdisk is approximately 170 $\mu$W (680 mW) for the pump, and 4 $\mu$w (400 $\mu$W) for the signal. To account for a slight temperature change in the microdisk as we swap the CW with pulsed light waves, we fine tune the wavelengths of the pump and signal by less than 0.05 nm so that they hit their respective slightly drifted resonance centers. Once good phase matching is achieved, the signal pulse overlapping a pump pulse will be deflected from the cavity via QZB, as shown in Fig. 5(b). The highest modulation extinction---defined as the ratio between the QZB-induced reduction in the transmission loss to the loss when the signal is at the cavity-resonance center without QZB---is 51.0\%, obtained when the signal is blue detuned from its resonance center by 0.01 nm. According to our simulation, this shifted optimal wavelength for the signal is likely due to the imperfect phase mismatching, where the generated SF light is slightly off a resonance center. Further blue detuning the signal reduces the extinction ratio to 16.3\% or less. In contrast, when the signal is red detuned, its depletion through the disk is enhanced by the pump, as shown in Fig. 5(c). These results show that the SFG driven by the pump effectively alters the signal's cavity resonance while shifting it to a shorter wavelength, similar to the ac Stark shift for a two-level atom \cite{harris1997electromagnetically}. In contrast to the thermally-induced resonance drift that affects all cavity lines, here the resonance altering applies to only a single resonant line satisfying the phase matching condition for SFG.
\begin{figure}[H]
  \centering

   \subfigure[]{
   \label{fig:subfig1.4.1:a} %% label for 1 subfigure
    \includegraphics[width=2.5in]{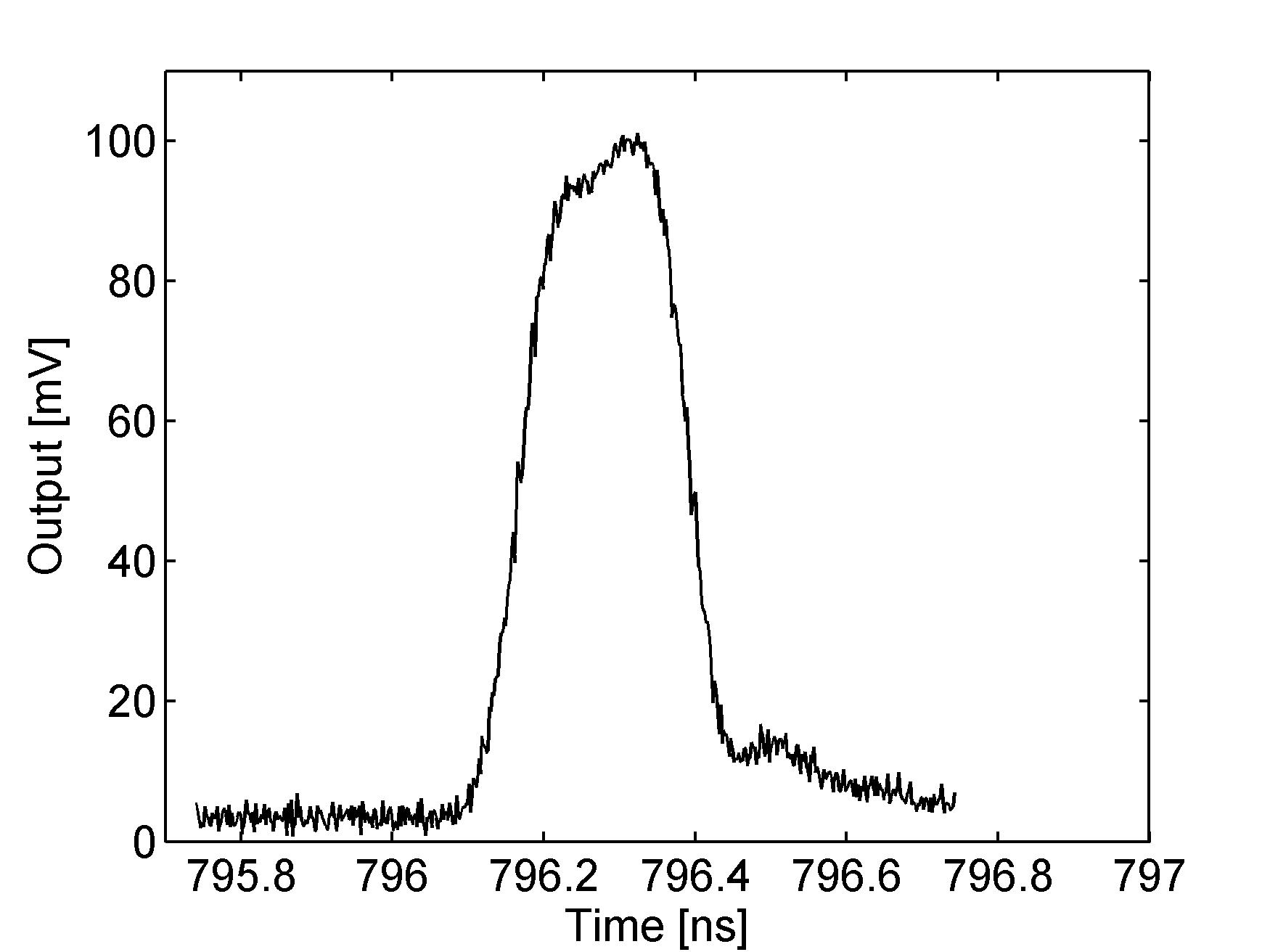}}

  \subfigure[]{
   \label{fig:subfig1.4.1:a} %% label for 1 subfigure
    \includegraphics[width=2.5in]{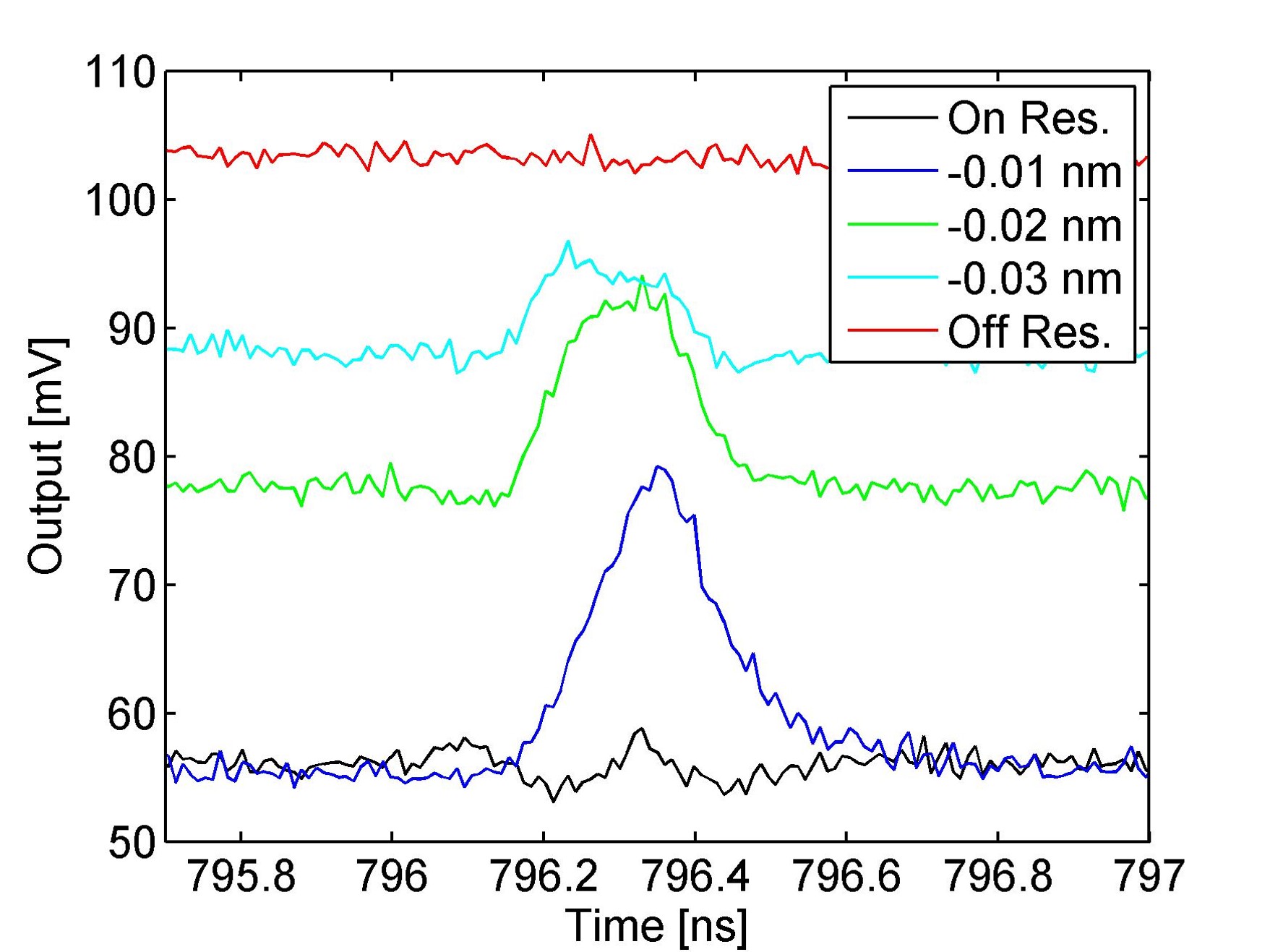}}
  \hspace{0.5in}
  \subfigure[]{
   \label{fig:subfig1.4.1:b} %% label for2 subfigure
    \includegraphics[width=2.5in]{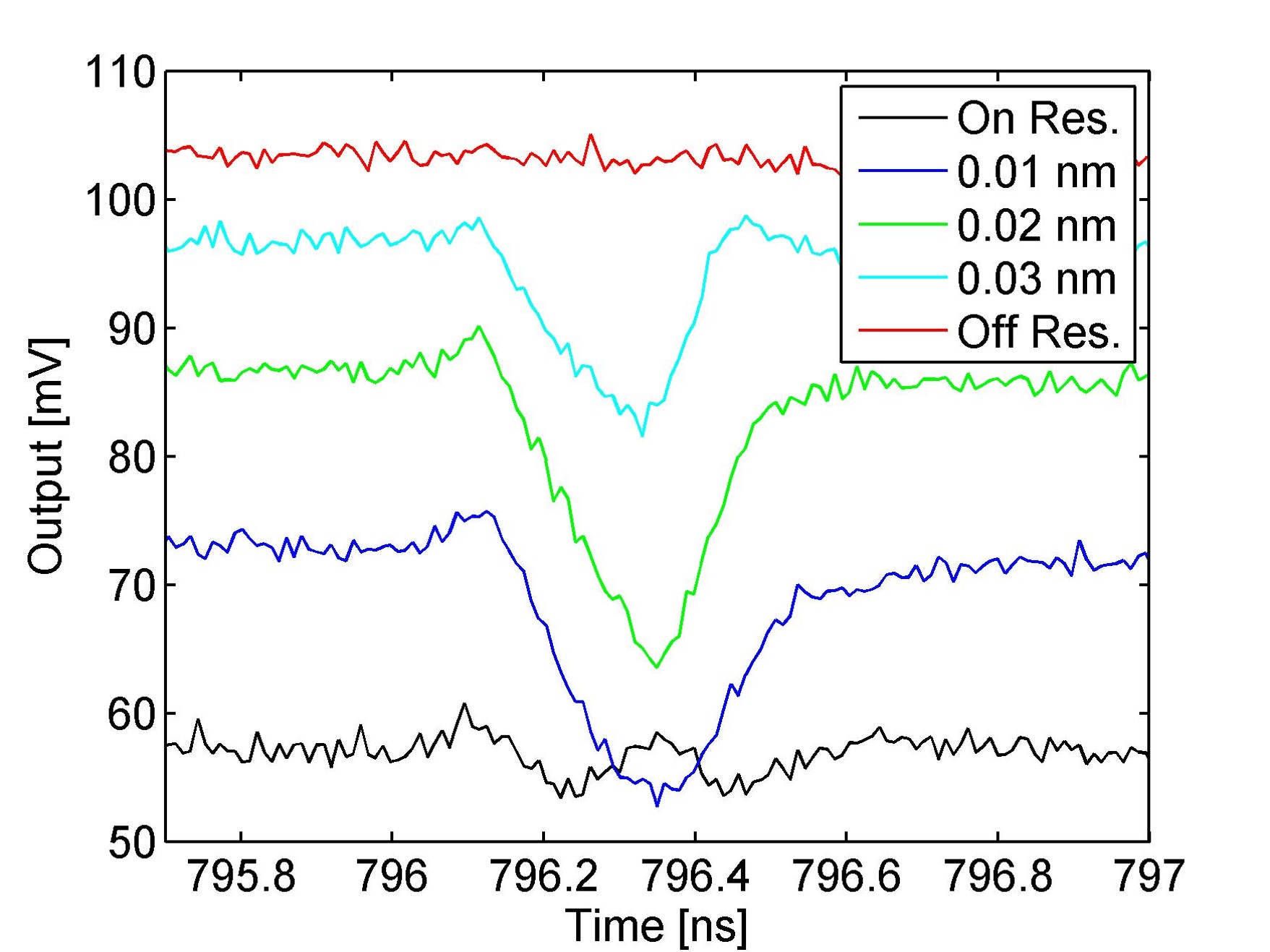}}

  \caption{(a) Temporal profile of the pump pulse with about 250 ps FWHM. (b) QZB-induced modulation on a quasi-CW signal as it is blue detuned from its resonance center. The resonant wavelength is 1536.790 nm for the pump and 1545.900 nm for the signal. (c) The depletion of the signal under the same pumping condition but when it is red detuned. The 3-dB baseline of the transmitted signal even when it is right on resonance in Figs. (b) and (c) is due to the under-coupling of the microdisk with the tapered fiber. }
  \label{fig:subfig1.4.1} %% label for entire figure
\end{figure}

Next, we examine QZB for operations on signals in pulses. In order to modulate the entire signal pulses, we use the same pump (250 ps FWHM) but narrower signal pulses with 200 ps FWHM. Furthermore, to distinguish QZB from any thermally-induced effect, we use different pulse repetition rates for the pump (1 MHz) and signal pulses (50 MHz), so that only one in every fifty signal pulses overlap with a pump pulse for QZB while the rest are unaffected. By keeping the same average power for the signal in the quasi-CW and pulsed cases, we are able to quickly swap them without disturbing the thermal equilibrium of the microdisk. The pump and signal are temporally aligned so that the peaks of the overlapping pulses are coincident within 10 ps. To contrast the signal transmission through the microdisk with and without QZB, we later use a tunable optical delay to temporally misplace the pulses. Figure 6 compares the transmission of the signal pulses as they are blue detuned from the resonance center. As seen, when there is no overlap between the signal and pump pulses, about 38.1\% of the input signal power is lost in the microdisk at the resonance center, which is a result of the cavity under-coupling. When the signal and pump are aligned, the lost power is reduced as the signal is deflected from the cavity by QZB. Specifically, at the resonance center, the signal loss with QZB becomes 29.3\%, which corresponds to a 23.0\% reduction. In contrast, when the signal is blue detuned by 0.01 nm, its transmittance loss becomes 14.4\% with the pump and 32.8\% without the pump, giving a 48.2\% modulation extinction, caused by the QZB. When the signal is further blue detuned to 0.02 nm, however, the extinction drops to 40.0\%. This interesting variation of QZB effectiveness is because the SFG is not perfectly phase matched in the current microdisk, as concord with our numerical simulation. All results are consistent with our measurement in the quasi-CW case.

%%  (b） delta 2.3 16.7 19
%% (c) 22.5 17.68
%%(d) 23.6 19.5

%(b-d) shows switching characteristics for a fixed pump wavelength and various signal wavelengths slightly detuned from cavity resonance (1536.790 nm). For the case that signal light has 0.01 nm of blue detuning, up to 20\% switching ratio can be achieved once the pump pulse is delayed to have the maximum overlapping with the signal pulse.

\begin{figure}
\includegraphics[width=5in]{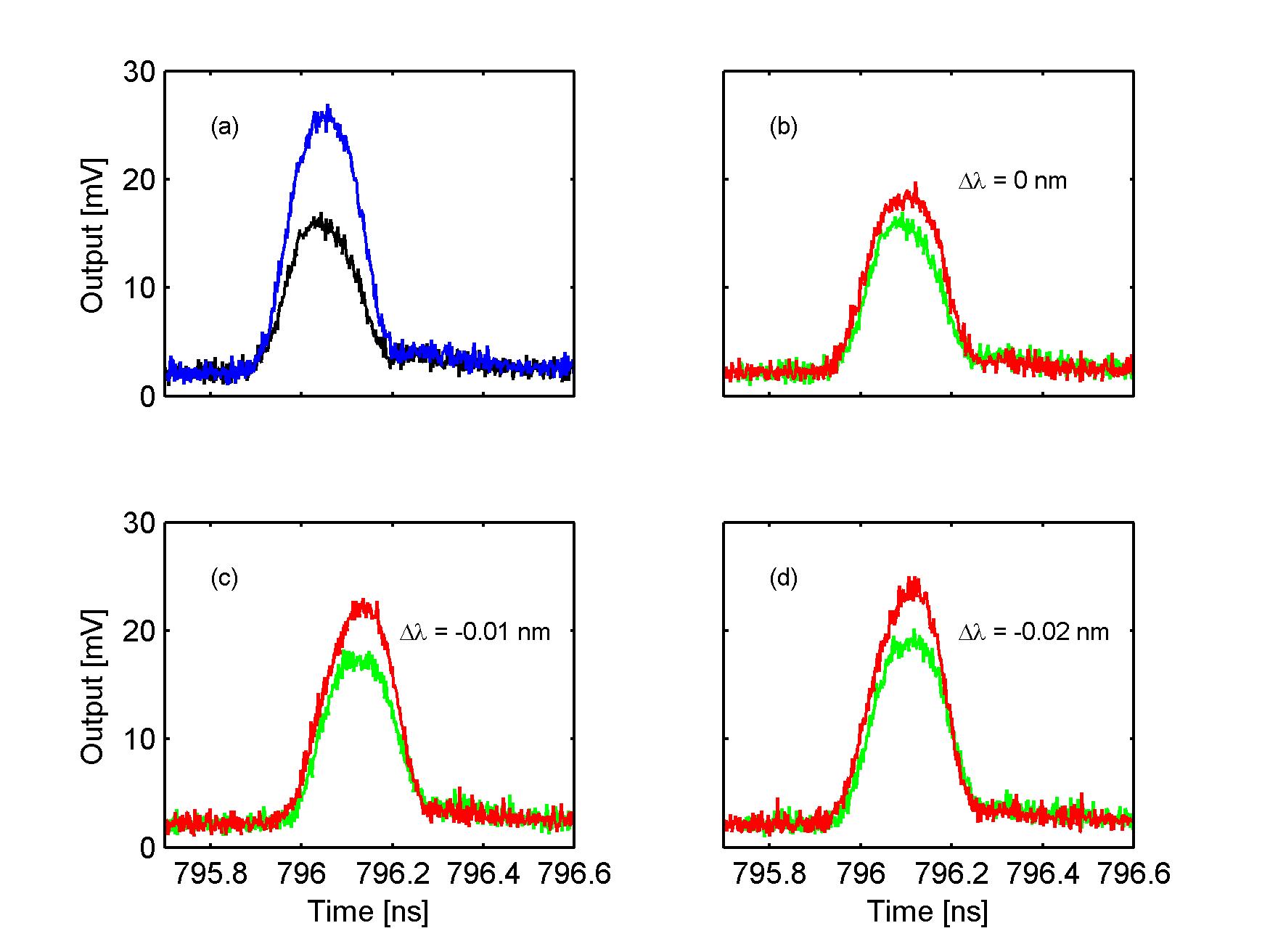} % this command will be ignored
  \centering
\caption{(a) The transmitted signal pulses in the on (black line) and off  (blue line) resonance case, when they are not temporally aligned with the pump. (b-d) The signal transmission when they are blue detuned from the resonance center (1536.790 nm) by $\Delta \lambda$ = 0, 0.01, 0.02 nm, respectively. In each figure, the green and red lines show the results when the signal pulse are aligned with with a pump pulse and completely displaced from the pump pulse by 300 ps, respectively.}
\end{figure}

\section{Single-photon operations}

We lastly assess the feasibility of our microdisk for quantum applications by quantifying its in-band photon noise level and demonstrating ``interaction-free'' modulation of single photons. Limited by our available filters (so as to attain the required $>$100 dB filtering extinction for the output signal photons), we select two non-adjacent fundamental quasi-TE WGM's for the pump and signal, at 1545.900 nm and 1564.750 nm, respectively. This wavelength selection unfortunately reduces the QZB efficiency by more than half as the SFG becomes less phase matched.

The experimental setup is shown in Fig. 4, where two CWDM filters are cascaded at the input to combine the pump and signal while rejecting their ASE noises, and two dense-wavelength-division-multiplexers (DWDM's) are cascaded at the output to pick the signal while rejecting the pump. Each DWDM filter is measured to have a 200-GHz FWHM bandwith and around 50 dB extinction. The resulting photons are counted using a commercial InGaAs single-photon detector (ID210, ID Quantique) with a 10.0\% quantum efficiency, and the measurement is averaged over 2 million signal pulses.

In our setup, the noise photons can be created extrinsically in the fiber connecting the cascaded CWDM and DWDM filters through spontaneous Raman scattering, as well as intrinsically in the LNOI microdisk via spontaneous Raman scattering and spontaneous parametric down conversion (SPDC) of the SH light created by the pump \cite{pelc2011long}. To quantify the intrinsic photon noise level, we first count the noise photons created in the fibers only, beforethe microdisk is coupled. Then, we bring the tapered fiber to the microdisk and count the photons again after verifying the fiber-microdisk coupling. The results are listed in Table 1, from which the intrinsic noise level is derived by comparing the photon counts with and without the microdisk. We note that the difference in the two detector gating rates is due to the detector dead time (10 $\mu$s) enforced after each photon detection, implemented to suppress the detector dark count.

\begin{table}
\centering
\caption{Measurement of photon noise.}
\medskip
\begin{tabular}{|c|c|c|c|c|}
\hline
Case & Gating Rate [KHz] & Detection Rate [KHz] & Input Power [uW] & Output Power [uW] \\

\hline
w/o disk & 939.4 ({$\pm$}0.1\%) & 6.65 ({$\pm$}1.0\%) & 770 & 509 \\
\hline
w/ disk & 935.6 ({$\pm$}0.8\%) & 7.12 ({$\pm$}1.0\%)& 776 & 332 \\
%\hline
%Note: The length of the connected fiber before (after) device is three (four) meters.\\

\hline
\hline
\end{tabular}
\end{table}

Specifically, we first use the results without the microdisk to derive the number of noise photons created in a unit length of fiber per Watt of pump power, $R_f$, through $ R_f L_1 P_1 \alpha_1 + R_f L_2 P_2 = D_1/\eta G_1$, where $L_1=3$ m and $L_2=4$ m are the fiber lengths before and after the microdisk, respectively, $P_1$ and $P_2$ are the input and output pump power, $G_1$ and $D_1$ are the detector gating rate and the photon counting rate. $\alpha_1$ is the linear loss in the tapered region, measured to be about $1.8$ dB, and $\eta$ is the total detection efficiency, which is 8.6\% for the current setup, due to transmission loss and quantum efficiency of the single photon detector. Then, with the disk, we apply $ R_f L_1 P_3 \alpha_2+ R_r+R_f L_2 P_4 =D_2/\eta G_2$ to calculate $R_r$, the rate of out-coupled noise photons created in the microdisk in the signal band. Here, $\alpha_2$ is the loss ($3.7$ dB) when coupled with the microdisk, $P_3$ and $P_4$ are the input and output power, respectively, and $G_2$ and $D_2$ are the gating and photon detection rates. This gives $R_r=9.3\times10^{-2}$ per pump pulse, for the current pump pulse FWHM of 250 ps and peak power of 680 mW. As the DWDM filters combined give a spectral bandwidth of 200 GHz (FWHM), the number of detected time-frequency modes is about 13. Thus, by pushing to single-mode detection, the noise photon probability can be reduced to $7.2\times10^{-3}$ per signal mode. By further detuning the signal and pump, the noise level can be suppressed to negligible \cite{pelc2011long}.

The relatively low intrinsic noise level of the LNOI microdisk---even for the present small detuning between the signal and pump---is due to its air suspended geometry \cite{jiang2014compact}, isolated phonon modes \cite{loudon1964raman}, and a relatively small Raman scattering cross-section \cite{pelc2011long,jin2001study}. Enhancing the sum-frequency generation using better mode overlapping or tailored pulse shapes will lower the pump power requirement and further suppress the background noise \cite{manurkar2016multidimensional}. We expect that even for the pump and signal in the telecom C-band, upon improved nanofabrication techniques for a higher cavity Q, the level of noise photon can reach or exceed the dark count level of a typical InGaAs single-photon detector, which is about $10^{-5}$ per pulse.
\begin{table}
\centering
\caption{Single photon modulation.}
\medskip
\begin{tabular}{|c|c|c|c|c|}
\hline
Detuning [nm] & No QZB & QZB & {$\Delta$} & Ext. \\
\hline
0 & 0.112 & 0.118 & 0.006 & 12.7\% \\
\hline
0.01 & 0.123  & 0.133 & 0.010 & 20.6\% \\
\hline
0.02 & 0.136 & 0.142 & 0.006 & 12.7\% \\
\hline
\hline
\end{tabular}

\end{table}

We next demonstrate the QZB modulation of single photons by using attenuated laser pulses, each containing 0.16 photons on average. Table 2 shows the measurement results as we blue detune the signal from the cavity resonance, after subtracting the total detection background at 7.12 KHz, which includes the detector dark counts and the in-band noise photons created by the pump in the microdisk and fibers. Without QZB, when the signal is on (off) resonance, the detection rate and the gating rate are 59.20 KHz (62.88 KHz) and 465.0 KHz (395.0 KHz), respectively, yielding a photon probability 0.112 (0.159) ({$\pm$}1.0\%). This amounts to a 29.6\% transmission loss at the resonance center. With QZB, the loss is reduced. The highest modulation extinction is 20.6{$\pm$}0.8\% obtained when the signal wavelength is blue detuned by 0.01 nm from resonance center, which is consistent with our preceding results using bright signal pulses. By using optimally phase-matched wavelengths for the signal and pump, we expect to improve the extinction by more than two folds with the current microdisk.

\section{Conclusion}
Using a nonlinear nanocavity, we have demonstrated quantum Zeno blockade on chip, where a light is modulated by another without them overlapping in the cavity in the asymptotic limit. This distinct "interaction-free" implementation overcomes several fundamental difficulties with quantum nonlinear optics, such as phase distortion, Raman scattering, and pulse distortion, which would otherwise prevent achieving high performance for few-photon logical applications. Toward this end, we show that single photons can be modulated with low excessive noise. Our results pave a chip-integration approach to some unvisited areas in quantum nonlinear optics, where exotic operations are promised in the decoherence-free sub-space. The next step is to improve the phase matching while replacing the current fiber taper with a waveguide coupler, based on which all-optical operations can be achieved at low error rate while cascaded on a single chip to realize complex functionalities. Further increasing the cavity quality factor can lead to strong photon photon interaction for "interaction-free" quantum logic gates with high fidelity.

\begin{methods}
\subsection{Microdisk fabrication procedure.}
The microdisk is fabricated on lithium niobate on insulator (LNOI, by NANOLN Inc.), which is a 500 nm lithium-niobate thin film bonded on a 3 $\mu$m silicon dioxide layer above lithium niobate substrate. To attain good phase matching, we use FEM simulation tools to identify the optimum thickness of LNOI thin film to be 200 nm. Thus, 300 nm LN thin film is etched away using Oxford Plasmalab 100 Inductively Coupled Plasma (ICP) system. After piranha cleaning and dehydration, the etched sample with remaining 200 nm LN thin film is submerged into cationic organic surface active agent (SurPass 3000+) solution to improve surface adhesion with resist prior to spinning 1000 nm thickness of negative electron beam  resist (ma-N 2410). Then the sample is pre-baked and patterned using EBL (Elionix ELS-G100, 100 keV). After development and post-bake, the sample is dry etched to obtain 200 nm thick microdisk by Argon milling process using the same ICP system. BOE (buffered oxide etcher, 6:1) is used to undercut the microdisk to form the air-suspended structure. The chip is diced into smaller pieces for the measurement.
\subsection{Optimization method for stabilizing the SFG.}
To achieve locking-free yet efficient and stable SFG, we firstly use an OSA to measure the transmission of a broadband light from the ASE. Then the resonance centers of those fundamental quasi-TE polarized WGM's (usually the smallest linewidth) in each FSR are selected and marked. We switch in the CW light at each resonant wavelength and continuously fine tune its wavelength to compensate the thermally-induced shift in cavity resonance. Upon the cavity reaching its equilibrium status, we confirm its actual resonant wavelength via the brightness and the mode distribution of the generated SH light. Repeating the steps to check all those modes, we could identify and choose the best two modes for sum frequency generation. Lastly, we couple two CW lights at the wavelengths of the selected modes and apply the same method to track the moving resonance centers. Once efficient SFG is observed and thermally stabilized, we quickly switch to quasi-CW signal and pulsed pump while retaining their wavelengths.
%\subsection{Method subsection.}

%Here is a description of a specific method used.  Note that the
%subsection heading ends with a full stop (period) and that the
%command is \verb|\subsection{}| not \verb|\subsection*{}|.

\end{methods}
%%%%%%%%%%%%%%%%%%%%%%%%%%%%%%%%%%%%%%%%%%%%%%%%%%%%%%%%%%%%%%%%%%%
%% Put the bibliography here, most people will use BiBTeX in
%% which case the environment below should be replaced with
%% the \bibliography{} command.

%\begin{thebibliography}{1}
 %\bibitem{dummy} Articles are restricted to 50 references, Letters
 %to 30.
 %\bibitem{dummyb} No compound references -- only one source per
 %reference.
 %\end{thebibliography}

%\bibliographystyle{naturemag}
%\bibliography{sample}

\begin{thebibliography}{10}
\expandafter\ifx\csname url\endcsname\relax
  \def\url#1{\texttt{#1}}\fi
\expandafter\ifx\csname urlprefix\endcsname\relax\def\urlprefix{URL }\fi
\providecommand{\bibinfo}[2]{#2}
\providecommand{\eprint}[2][]{\url{#2}}

\bibitem{misra1977zeno}
\bibinfo{author}{Misra, B.} \& \bibinfo{author}{Sudarshan, E.~G.}
\newblock \bibinfo{title}{The zeno’s paradox in quantum theory}.
\newblock \emph{\bibinfo{journal}{Journal of Mathematical Physics}}
  \textbf{\bibinfo{volume}{18}}, \bibinfo{pages}{756--763}
  (\bibinfo{year}{1977}).

\bibitem{peres1980zeno}
\bibinfo{author}{Peres, A.}
\newblock \bibinfo{title}{Zeno paradox in quantum theory}.
\newblock \emph{\bibinfo{journal}{American Journal of Physics}}
  \textbf{\bibinfo{volume}{48}}, \bibinfo{pages}{931--932}
  (\bibinfo{year}{1980}).

\bibitem{joos1984continuous}
\bibinfo{author}{Joos, E.}
\newblock \bibinfo{title}{Continuous measurement: Watchdog effect versus golden
  rule}.
\newblock \emph{\bibinfo{journal}{Physical Review D}}
  \textbf{\bibinfo{volume}{29}}, \bibinfo{pages}{1626} (\bibinfo{year}{1984}).

\bibitem{ItaHeiBol1990PRA}
\bibinfo{author}{Itano, W.~M.}, \bibinfo{author}{Heinzen, D.~J.},
  \bibinfo{author}{Bollinger, J.~J.} \& \bibinfo{author}{Wineland, D.~J.}
\newblock \bibinfo{title}{Quantum zeno effect}.
\newblock \emph{\bibinfo{journal}{Phys. Rev. A}} \textbf{\bibinfo{volume}{41}},
  \bibinfo{pages}{2295--2300} (\bibinfo{year}{1990}).

\bibitem{elitzur1993quantum}
\bibinfo{author}{Elitzur, A.~C.} \& \bibinfo{author}{Vaidman, L.}
\newblock \bibinfo{title}{Quantum mechanical interaction-free measurements}.
\newblock \emph{\bibinfo{journal}{Foundations of Physics}}
  \textbf{\bibinfo{volume}{23}}, \bibinfo{pages}{987--997}
  (\bibinfo{year}{1993}).

\bibitem{kwiat1995interaction}
\bibinfo{author}{Kwiat, P.}, \bibinfo{author}{Weinfurter, H.},
  \bibinfo{author}{Herzog, T.}, \bibinfo{author}{Zeilinger, A.} \&
  \bibinfo{author}{Kasevich, M.~A.}
\newblock \bibinfo{title}{Interaction-free measurement}.
\newblock \emph{\bibinfo{journal}{Physical Review Letters}}
  \textbf{\bibinfo{volume}{74}}, \bibinfo{pages}{4763} (\bibinfo{year}{1995}).

\bibitem{kwiat1996quantum}
\bibinfo{author}{Kwiat, P.}, \bibinfo{author}{Weinfurter, H.} \&
  \bibinfo{author}{Zeilinger, A.}
\newblock \bibinfo{title}{Quantum seeing in the dark}.
\newblock \emph{\bibinfo{journal}{Scientific American}}
  \textbf{\bibinfo{volume}{275}}, \bibinfo{pages}{72--78}
  (\bibinfo{year}{1996}).

\bibitem{tsegaye1998efficient}
\bibinfo{author}{Tsegaye, T.} \emph{et~al.}
\newblock \bibinfo{title}{Efficient interaction-free measurements in a
  high-finesse interferometer}.
\newblock \emph{\bibinfo{journal}{Physical Review A}}
  \textbf{\bibinfo{volume}{57}}, \bibinfo{pages}{3987} (\bibinfo{year}{1998}).

\bibitem{kwiat1999high}
\bibinfo{author}{Kwiat, P.~G.} \emph{et~al.}
\newblock \bibinfo{title}{High-efficiency quantum interrogation measurements
  via the quantum zeno effect}.
\newblock \emph{\bibinfo{journal}{Physical Review Letters}}
  \textbf{\bibinfo{volume}{83}}, \bibinfo{pages}{4725} (\bibinfo{year}{1999}).

\bibitem{hosten2006counterfactual}
\bibinfo{author}{Hosten, O.}, \bibinfo{author}{Rakher, M.~T.},
  \bibinfo{author}{Barreiro, J.~T.}, \bibinfo{author}{Peters, N.~A.} \&
  \bibinfo{author}{Kwiat, P.~G.}
\newblock \bibinfo{title}{Counterfactual quantum computation through quantum
  interrogation}.
\newblock \emph{\bibinfo{journal}{Nature}} \textbf{\bibinfo{volume}{439}},
  \bibinfo{pages}{949--952} (\bibinfo{year}{2006}).

\bibitem{huang2010interaction1}
\bibinfo{author}{Huang, Y.-P.} \& \bibinfo{author}{Kumar, P.}
\newblock \bibinfo{title}{Interaction-free all-optical switching in $\chi$ (2)
  microdisks for quantum applications}.
\newblock \emph{\bibinfo{journal}{Optics letters}}
  \textbf{\bibinfo{volume}{35}}, \bibinfo{pages}{2376--2378}
  (\bibinfo{year}{2010}).

\bibitem{huang2010interaction2}
\bibinfo{author}{Huang, Y.-P.}, \bibinfo{author}{Altepeter, J.~B.} \&
  \bibinfo{author}{Kumar, P.}
\newblock \bibinfo{title}{Interaction-free all-optical switching via the
  quantum zeno effect}.
\newblock \emph{\bibinfo{journal}{Physical Review A}}
  \textbf{\bibinfo{volume}{82}}, \bibinfo{pages}{063826}
  (\bibinfo{year}{2010}).

\bibitem{huang2012interaction}
\bibinfo{author}{Huang, Y.-P.} \& \bibinfo{author}{Kumar, P.}
\newblock \bibinfo{title}{Interaction-free quantum optical fredkin gates in
  $\chi^{(2)}$ microdisks}.
\newblock \emph{\bibinfo{journal}{IEEE Journal of Selected Topics in Quantum
  Electronics}} \textbf{\bibinfo{volume}{18}}, \bibinfo{pages}{600--611}
  (\bibinfo{year}{2012}).

\bibitem{mccusker2013experimental}
\bibinfo{author}{McCusker, K.~T.}, \bibinfo{author}{Huang, Y.-P.},
  \bibinfo{author}{Kowligy, A.~S.} \& \bibinfo{author}{Kumar, P.}
\newblock \bibinfo{title}{Experimental demonstration of interaction-free
  all-optical switching via the quantum zeno effect}.
\newblock \emph{\bibinfo{journal}{Physical review letters}}
  \textbf{\bibinfo{volume}{110}}, \bibinfo{pages}{240403}
  (\bibinfo{year}{2013}).

\bibitem{strekalov2014progress}
\bibinfo{author}{Strekalov, D.~V.}, \bibinfo{author}{Kowligy, A.~S.},
  \bibinfo{author}{Huang, Y.-P.} \& \bibinfo{author}{Kumar, P.}
\newblock \bibinfo{title}{Progress towards interaction-free all-optical
  devices}.
\newblock \emph{\bibinfo{journal}{Physical Review A}}
  \textbf{\bibinfo{volume}{89}}, \bibinfo{pages}{063820}
  (\bibinfo{year}{2014}).

\bibitem{jacobs2009all}
\bibinfo{author}{Jacobs, B.~C.} \& \bibinfo{author}{Franson, J.}
\newblock \bibinfo{title}{All-optical switching using the quantum zeno effect
  and two-photon absorption}.
\newblock \emph{\bibinfo{journal}{Physical Review A}}
  \textbf{\bibinfo{volume}{79}}, \bibinfo{pages}{063830}
  (\bibinfo{year}{2009}).

\bibitem{huang2012antibunched}
\bibinfo{author}{Huang, Y.-P.} \& \bibinfo{author}{Kumar, P.}
\newblock \bibinfo{title}{Antibunched emission of photon pairs via quantum zeno
  blockade}.
\newblock \emph{\bibinfo{journal}{Physical review letters}}
  \textbf{\bibinfo{volume}{108}}, \bibinfo{pages}{030502}
  (\bibinfo{year}{2012}).

\bibitem{birnbaum2005photon}
\bibinfo{author}{Birnbaum, K.~M.} \emph{et~al.}
\newblock \bibinfo{title}{Photon blockade in an optical cavity with one trapped
  atom}.
\newblock \emph{\bibinfo{journal}{Nature}} \textbf{\bibinfo{volume}{436}},
  \bibinfo{pages}{87--90} (\bibinfo{year}{2005}).

\bibitem{shapiro2006single}
\bibinfo{author}{Shapiro, J.~H.}
\newblock \bibinfo{title}{Single-photon kerr nonlinearities do not help quantum
  computation}.
\newblock \emph{\bibinfo{journal}{Physical Review A}}
  \textbf{\bibinfo{volume}{73}}, \bibinfo{pages}{062305}
  (\bibinfo{year}{2006}).

\bibitem{chang2014quantum}
\bibinfo{author}{Chang, D.~E.}, \bibinfo{author}{Vuleti{\'c}, V.} \&
  \bibinfo{author}{Lukin, M.~D.}
\newblock \bibinfo{title}{Quantum nonlinear optics-photon by photon}.
\newblock \emph{\bibinfo{journal}{Nature Photonics}}
  \textbf{\bibinfo{volume}{8}}, \bibinfo{pages}{685--694}
  (\bibinfo{year}{2014}).

\bibitem{sun2013photonic}
\bibinfo{author}{Sun, Y.-Z.}, \bibinfo{author}{Huang, Y.-P.} \&
  \bibinfo{author}{Kumar, P.}
\newblock \bibinfo{title}{Photonic nonlinearities via quantum zeno blockade}.
\newblock \emph{\bibinfo{journal}{Physical review letters}}
  \textbf{\bibinfo{volume}{110}}, \bibinfo{pages}{223901}
  (\bibinfo{year}{2013}).

\bibitem{fortsch2013versatile}
\bibinfo{author}{F{\"o}rtsch, M.} \emph{et~al.}
\newblock \bibinfo{title}{A versatile source of single photons for quantum
  information processing}.
\newblock \emph{\bibinfo{journal}{Nature Communications}}
  \textbf{\bibinfo{volume}{4}}, \bibinfo{pages}{1818} (\bibinfo{year}{2013}).

\bibitem{SecondGaAs}
\bibinfo{author}{Kuo, P.~S.}, \bibinfo{author}{Bravo-Abad, J.} \&
  \bibinfo{author}{Solomon, G.~S.}
\newblock \bibinfo{title}{Second-harmonic generation using -quasi-phasematching
  in a gaas whispering-gallery-mode microcavity}.
\newblock \emph{\bibinfo{journal}{Nature Communications}}
  \textbf{\bibinfo{volume}{5}}, \bibinfo{pages}{3109} (\bibinfo{year}{2014}).

\bibitem{moore2016efficient}
\bibinfo{author}{Moore, J.} \emph{et~al.}
\newblock \bibinfo{title}{Efficient second harmonic generation in lithium
  niobate on insulator}.
\newblock In \emph{\bibinfo{booktitle}{CLEO: Science and Innovations}},
  \bibinfo{pages}{STh3P--1} (\bibinfo{organization}{Optical Society of
  America}, \bibinfo{year}{2016}).

\bibitem{meisenheimer2015broadband}
\bibinfo{author}{Meisenheimer, S.-K.} \emph{et~al.}
\newblock \bibinfo{title}{Broadband infrared spectroscopy using optical
  parametric oscillation in a radially-poled whispering gallery resonator}.
\newblock \emph{\bibinfo{journal}{Optics express}}
  \textbf{\bibinfo{volume}{23}}, \bibinfo{pages}{24042--24047}
  (\bibinfo{year}{2015}).

\bibitem{harun2013theoretical}
\bibinfo{author}{Harun, S.}, \bibinfo{author}{Lim, K.}, \bibinfo{author}{Tio,
  C.}, \bibinfo{author}{Dimyati, K.} \& \bibinfo{author}{Ahmad, H.}
\newblock \bibinfo{title}{Theoretical analysis and fabrication of tapered
  fiber}.
\newblock \emph{\bibinfo{journal}{Optik-International Journal for Light and
  Electron Optics}} \textbf{\bibinfo{volume}{124}}, \bibinfo{pages}{538--543}
  (\bibinfo{year}{2013}).

\bibitem{harris1997electromagnetically}
\bibinfo{author}{Harris, S.~E.}
\newblock \bibinfo{title}{Electromagnetically induced transparency}.
\newblock \emph{\bibinfo{journal}{Physics Today}}
  \textbf{\bibinfo{volume}{50}}, \bibinfo{pages}{36--42}
  (\bibinfo{year}{1997}).

\bibitem{pelc2011long}
\bibinfo{author}{Pelc, J.~S.} \emph{et~al.}
\newblock \bibinfo{title}{Long-wavelength-pumped upconversion single-photon
  detector at 1550 nm: performance and noise analysis}.
\newblock \emph{\bibinfo{journal}{Optics express}}
  \textbf{\bibinfo{volume}{19}}, \bibinfo{pages}{21445--21456}
  (\bibinfo{year}{2011}).

\bibitem{jiang2014compact}
\bibinfo{author}{Jiang, W.~C.}, \bibinfo{author}{Zhang, J.} \&
  \bibinfo{author}{Lin, Q.}
\newblock \bibinfo{title}{Compact suspended silicon microring resonators with
  ultrahigh quality}.
\newblock \emph{\bibinfo{journal}{Optics express}}
  \textbf{\bibinfo{volume}{22}}, \bibinfo{pages}{1187--1192}
  (\bibinfo{year}{2014}).

\bibitem{loudon1964raman}
\bibinfo{author}{Loudon, R.}
\newblock \bibinfo{title}{The raman effect in crystals}.
\newblock \emph{\bibinfo{journal}{Advances in Physics}}
  \textbf{\bibinfo{volume}{13}}, \bibinfo{pages}{423--482}
  (\bibinfo{year}{1964}).

\bibitem{jin2001study}
\bibinfo{author}{Jin, C.} \& \bibinfo{author}{Zhao, C.}
\newblock \bibinfo{title}{A study on the raman spectrum of the linbo3-doped mg
  crystal at low temperature}.
\newblock \emph{\bibinfo{journal}{Spectroscopy Letters}}
  \textbf{\bibinfo{volume}{34}}, \bibinfo{pages}{437--442}
  (\bibinfo{year}{2001}).

\bibitem{manurkar2016multidimensional}
\bibinfo{author}{Manurkar, P.} \emph{et~al.}
\newblock \bibinfo{title}{Multidimensional mode-separable frequency conversion
  for high-speed quantum communication}.
\newblock \emph{\bibinfo{journal}{Optica}} \textbf{\bibinfo{volume}{3}},
  \bibinfo{pages}{1300--1307} (\bibinfo{year}{2016}).

\end{thebibliography}

%% Here is the endmatter stuff: Supplementary Info, etc.
%% Use \item's to separate, default label is "Acknowledgements"

\begin{addendum}
 \item This research was support in part by the National Science Foundation (Award No. ECCS-1521424). Fabrication was performed at CUNY Advanced Science Research Center Nano Fabrication Facility. We also acknowledge Stefan Strauf for assistance in setting up the fiber tapering system and helpful discussions.

\item[Contribution] J.-Y.C. fabricated the chip device with help from M.L. Y.M.S and Z.-T.Z made the fiber taper. J.-Y.C., Y.M.S., and Z.-T.Z. performed the experiment. Y.-P.H conceived and supervised the project. All contributed to writing the paper.

 \item[Competing Interests] The authors declare that they have no
competing financial interests.
 \item[Correspondence] Correspondence and requests for materials
should be addressed to yuping huang ~(email: yuping.huang@stevens.edu).
\end{addendum}

%%
%% TABLES
%%
%% If there are any tables, put them here.
%%

%\begin{table}
%\centering
%\caption{This is a table with scientific results.}
%\medskip
%\begin{tabular}{ccccc}
%\hline
%1 & 2 & 3 & 4 & 5\\
%\hline
%aaa & bbb & ccc & ddd & eee\\
%aaaa & bbbb & cccc & dddd & eeee\\
%aaaaa & bbbbb & ccccc & ddddd & eeeee\\
%aaaaaa & bbbbbb & cccccc & dddddd & eeeeee\\
%1.000 & 2.000 & 3.000 & 4.000 & 5.000\\
%\hline
%\end{tabular}
%\end{table}

\end{document}